# A perspective on the Healthgrid initiative

V. Breton[1], A.E. Solomonides[2], R.H. McClatchey[2]

1 LPC, CNRS-IN2P3, Université Blaise Pascal, Campus des Cézeaux, 63177 Aubière Cedex, France

2 CCCS, UWE Bristol, Frenchay Campus, Bristol BS16 1QY, UK

breton@clermont.in2p3.fr, {tony.solomonides / richard.mcclatchey}@uwe.ac.uk

## Abstract

*This paper presents a perspective on the Healthgrid initiative which involves European projects deploying pioneering applications of grid technology in the health sector. In the last couple of years, several grid projects have been funded on health related issues at national and European levels. A crucial issue is to maximize their cross fertilization in the context of an environment where data of medical interest can be stored and made easily available to the different actors in healthcare, physicians, healthcare centres and administrations, and of course the citizens. The Healthgrid initiative, represented by the Healthgrid association (http://www.healthgrid.org), was initiated to bring the necessary long term continuity, to reinforce and promote awareness of the possibilities and advantages linked to the deployment of GRID technologies in health. Technologies to address the specific requirements for medical applications are under development. Results from the DataGrid and other projects are given as examples of early applications.*

## 1. Grid impact for healthcare: an example

Last summer, about 10,000 elderly people died in one European country because of unusually prolonged, severely hot weather. For two weeks, the overall increase in mortality rate in hospitals and healthcare centres remained unnoticed. As a means to better handle this kind of situation, a proposed strategy is to set up a monitoring service recording daily on a central repository the number of casualties in each healthcare centre. With the present telemedicine tools, such a monitoring service requires an operator in each healthcare centre to submit the information to the central repository and an operator to validate the information provided. In case of emergency, for instance if the monitoring service identifies an abnormal increase of the mortality rate, experts have to be called to analyze the information available at the central repository. If they want additional information, they need to ask for it from the operators in each healthcare centre. This extra request may introduce further delays and extra work on health professionals who are already overworked.

With the deployment of grid technology, such a monitoring service would require much less manpower. Indeed, grid technology already allows users today to access in a secure way to data stored on distant grid nodes. Instead of having one operator in each centre in charge of transmitting information daily to the central repository, the information on the number of casualties is stored locally on a database which is accessible by the central repository. In case of emergency, the experts can access deeper into the healthcare centre database to inquire about particular cases. The picture is completely different: patient medical files stay in healthcare centres and the central monitoring service picks up only what is needed for its task.

## 2. Vision for a grid for health

Grid technology has been identified as one of the key technologies to enable the 'European Research Area'. The impact of this concept is expected to reach far beyond eScience, to eBusiness, eGovernment, and eHealth. However, a major challenge is to take the technology out of the laboratory to the citizen. A Healthgrid is an environment where data of medical interest can be stored and made easily available to the different actors of healthcare, physicians, healthcare centres and administrations, and of course citizens. But, such an environment has to offer all guarantees in terms of security, respect for ethics and observance of regulations. Moreover, the association of post-genomic information and medical data in such an environment opens up the possibility of individualized healthcare. While considering the deployment of life sciences applications, most present grid projects do not address the specificities of an e-infrastructure for health, for instance the deployment of grid nodes in clinical centres and in healthcare administrations, the connection of individual physicians to the grid and the strict regulations ruling the access to personal data.

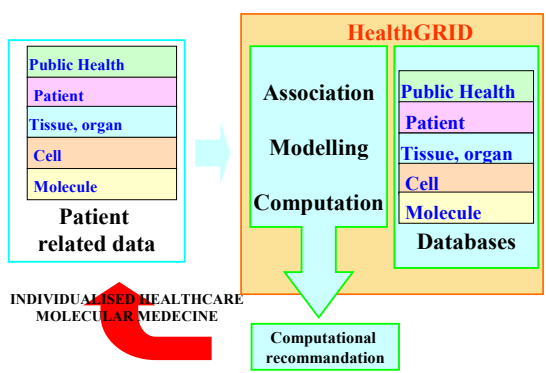

Technology to address these requirements in a grid environment is under development and pioneering work is underway in the application of Grid technologies to the health area.

In the last couple of years, several grid projects have been funded on health related issues at national and European levels. These projects have a limited lifetime, from 3 to 5 years, and a crucial issue is to maximize their cross fertilization. Indeed, the Healthgrid is a long term vision that needs to build on the contribution of all projects. The Healthgrid initiative, represented by the Healthgrid association (http://www.healthgrid.org), was initiated to bring the necessary long term continuity. Its goal is to collaborate with projects on the following activities:

- Identification of potential business models for medical Grid applications.
- Feedback to the Grid development community on the requirements of the pilot applications deployed by the European projects.
- Development of a systematic picture of the broad and specific requirements of physicians and other health workers when interacting with Grid applications.
- Dialogue with clinicians and those involved in medical research and Grid development to determine potential pilots.
- Interaction with clinicians and researchers to gain feedback from the pilots.
- Interaction with all relevant parties concerning legal and ethical issues identified by the pilots.
- Dissemination to the wider biomedical community on the outcome of the pilots.
- Interaction and exchange of results with similar groups worldwide.
- The formulation and specification of potential new applications in conjunction with the end user communities.

## 3. Requirements

The example used to introduce this paper illustrates the potential impact of Grid technology for health. To deploy the monitoring service described, patients' medical data would have to be stored in a local database in each healthcare centre. These databases would have to be federated and would share the same logical model. Secure access to data would have to be granted: only views of local data would be available to external services and patient data would have to be anonymized.

These requirements need to be fed back to the community developing middleware services. The Healthgrid initiative is actively involved in the definition of requirements relevant to the usage of grids for health. As an example of a first list of requirements, the one produced by DataGrid [1] can be given.

### 3.1 Data related requirements
- Provide access to biology and medical image data from various existing databases.
- Support and improve existing databases import/export facilities.
- Provide transparent access to data from the user point of view, without knowledge of their actual location.

- Update databases while applications are still running on their own data versions.

## 3.2 Security related requirements

- Allow anonymous and private login for access to public and private databases.
- Guarantee the privacy of medical information.
- Provide an encryption of sensitive message.
- Fulfill all legal requirements in terms of data encryption and protection of patient privacy.
- Allow maintenance of security policies without the need to modify applications.

## 3.3 Administrative requirements

- Provide batch, prioritized batch, and interactive computing.
- Provide 'Virtual Grids', i.e. the ability to define sub-grids with a restricted access to data and computing power
- Provide 'Virtual Grids' with customizable login policies

## 3.4 Network requirements

- Allow batch computing on huge datasets.
- Allow fast processing applications which transfer small set of images between different sites: storage site, processing site and physician site.
- Allow interactive access to small data sets, like image slices and model geometry.

## 3.5 Job related requirements

- Allow the user to run jobs "transparently", without the knowledge of the underlying scheduling mechanism and resource availability
- Manage jobs priorities.
- Deal with critical jobs: if no resources are available, jobs with lowest priority will be interrupted to allow execution of critical jobs.
- Allow to chain batch jobs into pipelines.
- Provide a fault-tolerant infrastructure with fault detection, logging and recovery.
- Notify a user when a job fails for a Grid-independent reason.
- Provide an interactive mode for some applications.
- Allow parallel jobs.
- Provide an MPI-like interface.
- Allow message passing inside a local farm and also at the Grid level.
- Offer job monitoring, such as query job status, cancel queuing or running job.
- Provide understanding of job failure; provide log files.

## 4. Early results of application deployment

Today, grid technology is still under development and standards are just emerging. In this chapter, we present results of pilot applications deployed on the DataGrid and other projects. These applications provide basic common services (web portals, computing resources). A pioneer action in the area of medical development is also described. Further examples from other European projects can be found on Healthgrid web site.

### 4.1. Parallelization of Monte-Carlo simulations for nuclear medical imaging (DataGrid)

Monte Carlo simulations are increasingly used in medical physics. In nuclear medical imaging these simulations are used to model imaging systems and to develop and assess tomographic reconstruction algorithms and correction methods for improved image quantization. In radiotherapy-brachytherapy the goal is to evaluate accurately the dosimetry in complex phantoms and at interfaces of tissue, where analytic calculations have shown some limits. But the main drawback of Monte Carlo simulations is long compute time. Grids are expected to reduce this time by parallelizing a simulation on geographically distributed processors. The method used for parallelization of the Random Number Generator is to split the long number used by the sequential simulation. Once the partitioning is done, a software application allows the user to generate automatically the files describing each simulation part and executes them on the DataGrid test bed using an API [2]. Different tests were done in order to show, first, the reliability of the physical results obtained by concatenation of parallelized output data and secondly the time gained for job execution [3].

## 4.2 Bioinformatics portals

Web portals offering services to biologists in genomics and post-genomics (proteomics, metabolomics, etc.) have to handle exponential growth in databases. As the volume of data expands, comparative analysis of genome and protein sequences becomes increasingly CPU intensive. Bioinformatics algorithms deployed on the DataGrid test bed are available on the *GPS@* portal (Institut de Biologie et Chimie des Protéines) for protein secondary structure analysis and on the *PhyloJava* portal (Laboratoire de Biologie Biométrie Evolutive) for phylogenetics. Up to 1000 jobs were submitted in parallel on the DataGrid test bed with observed reduction of computing time up to a factor of 10.

## 4.3. Humanitarian medical development

The training of local clinicians is the best way to raise the standard of medical knowledge in developing countries. This requires transferring skills, techniques and resources. Grid technology opens new perspectives for preparation and follow-up of medical missions in developing countries as well as support to local medical centres in terms of teleconsulting, telediagnosis, patient follow-up and e-learning. To meet requirements of a development project of the French NPO *Chain of Hope* in China, an initial protocol was established for describing the patients' pathologies and their pre- and post-surgery states through a web interface in a language-independent way. This protocol was evaluated by French and Chinese clinicians during medical missions in the fall 2003. The first sets of medical patients recorded in the databases will be used to evaluate grid implementation of services and to deploy a grid-based federation of databases. Such a federation of databases allows medical data to be kept distributed in the hospitals behind firewalls. Views of the data will be granted according to individual access rights through secured networks.

## 5. Breast Cancer and Mammography

We further illustrate the range of European projects with a brief overview of three projects on mammography, concentrating mainly on the EU project MammoGrid. Breast cancer as a medical condition, and mammograms as images, exhibit many dimensions of variability across the population. Likewise, the way diagnostic systems are used and maintained by clinicians varies between imaging centres and breast screening programmes, as does the appearance of the mammograms generated. A distributed database that reflects the spread of pathologies across the population would be an invaluable tool for the epidemiologist, while understanding of the variation in image acquisition protocols is essential to a radiologist or radiographer (radiologic technician) in a screening programme. Exploiting emerging grid technology, the aim of the MammoGrid project is to develop a Europe-wide database of mammograms that will be used to investigate a set of important healthcare applications and to explore the potential of the grid to support effective co-working between healthcare professionals. In particular, the project aims to prove that grids infrastructures can be practically used for collaborative medical image analysis. This leads to several technical issues, including the standardization of mammograms, design of an appropriate clinical workstation and distribution of data, images and clinician queries across a grid-based database while respecting patient confidentiality and security protocols. The MammoGrid project aims to prove the viability of the grid by harnessing its power to enable radiologists from geographically dispersed hospitals to share standardized mammograms, to compare diagnoses (with and without computer aided detection of tumours) and to perform sophisticated epidemiological studies across national boundaries. MammoGrid has defined an imaging workstation architecture, an information infrastructure to connect radiologists across a Grid, and a DICOM-compliant object model residing in multiple, distributed data stores, currently in Italy and the UK. a number of relevant technologies that are being harnessed together in the MammoGrid project to provide a distributed infrastructure to support radiologists in their work. These include Mirada Solutions' Standard Mammogram FormTM; aspects of the CERN/CMS CRISTAL kernel; AliEn, a lightweight Grid developed for CERN/ALICE; and the computer-aided detection (CADe) software developed in the *CALMA* project in Italy. The project thus addresses requirements arising from:

- Image variability, due to differences in acquisition processes and to differences in the software packages (and underlying algorithms) used in their processing.

- Population variability, which causes regional differences affecting the various criteria used for the screening and treatment of breast cancer.

- Support for radiologists, in the form of tele-collaboration, second opinion, training, quality control of images and a growing evidence-base.

In practical terms, the project will:

- evaluate current Grids technologies and determine the requirements for Grid-compliance in a pan-European mammography database;
- implement a prototype MammoGrid database, using novel Grid-compliant and federated-database technologies that will provide improved access to distributed data;
- deploy versions of a standardization system that enables comparison of mammograms in terms of tissue properties independently of scanner settings, and to explore its place in the context of medical image formats; and
- use the annotated information and the images in the database to benchmark the performance of the prototype system.

A related two-year project, eDiamond [5], funded by the UK government is using similar radiological technologies but a radically different approach to the Grid (based on IBM architectures) and is focussed on training requirements. The Italian INFN project *GP-CALMA* (Grid Project CALMA) [6] has focussed on a Grid implementation of tumour and microcalcification detection algorithms to provide clinicians with a working mammogram examination tool.

# 6. An emerging pilot project: a grid to address an orphan disease

Another area where grid technology offers promising perspectives for health is drug discovery. This chapter presents the potential interest of a grid dedicated to research and development on a rare disease.

## 6.1 The crisis of neglected diseases
There is presently a crisis in research and development for drugs for neglected diseases. Infectious diseases kill 14 million people each year, more than ninety percent of whom are in the developing world. Access to treatment for these diseases is problematic because the medicines are unaffordable, some have become ineffective due to resistance, and others are not appropriately adapted to specific local conditions and constraints. Despite the enormous burden of disease, drug discovery and development targeted at infectious and parasitic diseases in poor countries has virtually ground to a standstill, so that these diseases are de facto neglected. Of the 1393 new drugs approved between 1975 and 1999, less than 1% was specifically for tropical diseases. Only a small percentage of global expenditure on health research and development, estimated at US$50-60 billion annually, is devoted to the development of such medicines. At the same time, the efficacy of existing treatments has fallen, due mainly to emerging drug resistance. [7]

## 6.2 The grid impact
The unavailability of appropriate drugs to treat neglected diseases is among other factors a result of the lack of ongoing R&D into these diseases. While basic research often takes place in university or government labs, development is almost exclusively done by the pharmaceutical industry, and the most significant gap is in the translation of basic research through to drug development from the public to the private sector. Another critical point is the launching of clinical trials for promising candidate drugs.
Producing more drugs for neglected diseases requires building a focussed, disease-specific R&D agenda including short-, mid- and long-term projects. It requires also a public-private partnership through collaborations that aim to improve access to drugs and stimulate discovery of easy-to-use, affordable, effective drugs.

The grid should gather:
- drug designers to identify new drugs
- healthcare centres involved in clinical tests
- healthcare centres collecting patent information
- structures involved in distributing existing treatments (healthcare administrations, non profit organizations,…)
- IT technology developers
- computing centres
- biomedical laboratories searching for vaccines, working on the genomes of the virus and/or the parasite and/or the parasite vector

The grid will be used as a tool for:
- search of new drug targets through post-genomics requiring data management and computing

- massive docking to search for new drugs requiring high performance computing and data storage
- handling of clinical tests and patent data requiring data storage and management
- overseeing the distribution of the existing drugs requiring data storage and management

A grid dedicated to research and development on a given disease should provide the following services:

- large computing resources for search for new targets and virtual docking
- large resources for storage of post genomics and virtual docking data output
- grid portal to access post genomics and virtual docking data
- grid portal to access medical information (clinical tests, drug distribution,…)
- a collaboration environment for the participating partners. No one entity can have an impact on all R&D aspects involved in addressing one disease.

Such a project would form the core of a community pioneering the use of grid-enabled medical applications. The choice of a neglected disease should help reduce participants' reluctance to communicate information. However, the issue of Intellectual Property must be addressed in thorough detail.

## 7. Conclusion

This article has illustrated different aspects relevant to the deployment of grid technology for health, from drug discovery to telediagnosis. Technologies to address the requirements for deployment of medical applications in a grid environment are under development and a pioneering work is under way in the application of Grid technologies to the health area in several European projects. Results from the DataGrid project are given as examples.

Today the first projects deploying biomedical applications have moved into their final phase (DataGrid), some which had only started last year have preliminary results (GEMSS, MammoGrid, etc.), while a new generation of large-scale grid infrastructure projects are just emerging (EGEE).

## 8. Acknowledgements

The authors acknowledge the ideas and contributions of their colleagues and collaborators in the Healthgrid Association and in the MammoGrid project.